\documentclass[twoside,reqno]{HERON}
\usepackage{epsfig,cite,colordvi}
\usepackage{graphicx}
\usepackage{url}
\usepackage{amssymb,amsmath,amscd,epsf}
\usepackage{times}
\usepackage{makeidx}
\begin{document}

\title{Testing nuclear models via neutrino scattering}
 

\runningheads{Testing nuclear models via neutrino scattering}{M.B. Barbaro {\it et al.}} 

\begin{start}

\author{M.B. Barbaro}{1},
\coauthor{C. Albertus}{2},
\coauthor{J.E. Amaro}{2},
\coauthor{A.N. Antonov}{3}, 
\coauthor{J.A. Caballero}{4}, 
\coauthor{T.W. Donnelly}{5}, 
\coauthor{R. Gonzalez-Jimenez}{4}, 
\coauthor{M.V. Ivanov}{3}, 
\coauthor{E. Moya de Guerra}{6}, 
\coauthor{G.D. Megias}{4}, 
\coauthor{I. Ruiz Simo}{2}, 
\coauthor{J.M. Udias}{6} 

\index{Barbaro, M.B.}
\index{Albertus, C.}
\index{Amaro, J.E.}
\index{Antonov, A.N.}
\index{Caballero, J.A.}
\index{Gonzalez-Jimenez, R.}
\index{Ivanov, M.V.}
\index{Megias, G.D.}
\index{Ruiz-Simo, I.}
\index{Udias, J.M.}

\address{University of Turin and INFN, 10128 Turin, Italy}{1}

\address{University of Granada, 18071 Granada, Spain}{2}

\address{INRNE, 1784 Sofia, Bulgaria}{3}

\address{University of Seville, 41080 Seville, Spain}{4} 

\address{M.I.T., 02139 Cambridge, MA, USA}{5}

\address{Universidad Complutense de Madrid, 28040 Madrid, Spain}{6}

\begin{Abstract}
Recent progresses on the relativistic modeling of neutrino-nucleus reactions are presented and the results are compared with high precision experimental data in a wide energy range.
\end{Abstract}

\end{start}

\section{Introduction}

Experimental knowledge of neutrino-nucleus cross sections has reached unprecedented precision in recent years, offering new opportunities to test models for the weak nuclear response. 
Several ongoing experiments (MiniBooNE~\cite{AguilarArevalo:2008qa}, MINER$\nu$A~\cite{Aliaga:2013uqz}, T2K~\cite{Abe:2011ks}, ArgoNeuT~\cite{Anderson:2012vc}) aim at the precise measurement of neutrino properties: masses and hierarchy, CP violation parameters and mixing angles. In order to get significant statistics, these experiments use complex nuclear targets - typically Carbon, Oxygen and Argon - and their analyses strongly rely on the modeling of nuclear effects, which are one of the main sources of uncertainty.

Ideally, neutrino scattering could provide richer information about the lepton-hadron reaction mechanism and the nuclear dynamics than electron scattering, giving access not only to the vector but also to the axial response. 
However, monochromatic neutrino beams are not available and all observables have to be folded with the experimental neutrino flux, which makes electrons more efficient probes of the vector nuclear response than neutrinos. 
For example, in inclusive electron scattering - where only the outgoing electron is detected in the final state - it is possible to disentangle the quasielastic (QE) contribution, corresponding to single-nucleon knockout,
from processes involving two or more nucleon knockout, since they occur at different transferred energies. 
Such separation is not possible in inclusive neutrino scattering, as the energy transfer is not precisely known. Therefore a reliable modeling of different nuclear processes becomes essential for a proper analysis of the experimental data.

All the above mentioned experiments involve neutrinos with energy around 1 GeV or more.
At these kinematics the dominant contributions to the cross section are quasielastic scattering and pion production, the latter occurring mainly through $\Delta$-resonance excitation. 
In this regime models based on a Lorentz covariant nuclear tensor, involving relativistic hadronic current operators and  wave functions, are preferable to traditional non-relativistic approaches.

 The simplest relativistic nuclear model is the Relativistic Fermi Gas (RFG), where relativistic effects can be treated exactly, but nucleons are considered free and correlated only by the Pauli principle.
It is well-known from electron scattering that effects which go beyond the RFG, such as nucleon-nucleon (NN) correlations and final state interactions (FSI), significantly affect the nuclear response. These can be accounted for, {\it e.g.}, in the Relativistic Mean Field (RMF) model, based on the solution of the Dirac equation in presence of strong scalar and vector potentials for both the initial and final state~\cite{Amaro:2011qb}, or in the Relativistic Green's Function (RGF) model, based on the use of a complex optical potential to describe FSI~\cite{Giusti:2014xfa,Meucci:2014bva}. The predictions of the two models have been compared and shown to give very similar  results for electron scattering but non-negligible differences for neutrino scattering, depending on the kinematics~\cite{MeucciPRL}.
 
An alternative method of accounting for nuclear effects in neutrino scattering in a model-independent way consists in extracting information on the many-body dynamics from electron scattering data. This approach leads to the so-called {\it superscaling} approximation (``SuSA''), which will be briefly reviewed in Section 2, together with the latest improvements to the model. 
In Section 3 the contribution of relativistic two-body currents, which go beyond superscaling,  will be discussed. In Section 4 we shall summarize and outline some future developments.

\section{Superscaling}

The SuSA model is a phenomenological approach based on the superscaling analysis of inclusive electron scattering data performed by Donnelly and Sick in Ref.~\cite{Donnelly:1999sw} for the quasielastic region and extended to non-quasielastic scattering in Refs.~\cite{Maieron:2001it,Barbaro:2003ie}.
 
In Ref.~\cite{Amaro:2004bs}, Amaro {\it et al.} proposed to use superscaling in order to predict charged-current (CC) neutrino scattering cross sections in the quasielastic and $\Delta$ resonance regions. The approach was later extended to the case of neutral-current (NC) scattering~\cite{Amaro:2006pr}.
 
The basic assumption of the SuSA model is that the cross section for inclusive lepton (electron or neutrino) scattering off a nucleus can be factorized into a single-nucleon function, which contains the appropriate kinematic factors and the elementary vertex ($B^* NN$ for QE scattering, $B^* N\Delta$ for resonant $\pi$-production, and so on for higher excitations - $B^*$ being the virtual boson, $\gamma$, $W^\pm$ or $Z^0$), times a function, the {\it scaling function} $F(\psi)$, depending on one single {\it scaling variable} $\psi(q,\omega)$ instead of $q$ (momentum transfer) and $\omega$ (energy transfer) separately. The exact definition of the scaling variable and dividing factors can be found {\it e.g.} in Ref.~\cite{Barbaro:1998gu}.
This is a good approximation as long as the probe interacts with the complex system (the nucleus) by transferring energy and momentum to the individual constituents (the nucleons). 
The idea is conceptually similar to the well-known Bjorken scaling, where the complex system is the nucleon and the constituents are the partons, with the important differences that nucleons, unlike partons, are not pointlike nor asymptotically free.
 
Scaling in nuclei is expected to be realized at high enough values of $q$ (larger than roughly 400 MeV/c), where collective effects are not present, and in absence of two-body currents, associated to the interaction of the probe with a pair of correlated nucleons. Moreover, if the function $F$ scales with the nuclear species as the inverse Fermi momentum, $1/k_F$, the phenomenon is called {\it superscaling} and $f(\psi)=k_F F(\psi)$ is the superscaling function, independent of the specific nucleus. This property allows to easily apply the model to any nucleus. 

Superscaling in the quasielastic region has been shown to be fulfilled with good accuracy by the longitudinal $(e,e')$ data, while it is violated in the transverse channel, where processes different from single-nucleon ejection, such as $\Delta$-resonance and multi-nucleon excitations, come into play. These contributions have to be added to the SuSA model to get a full description of the data.

\begin{figure}[h]
\centering
\includegraphics[scale=0.2,angle=270]{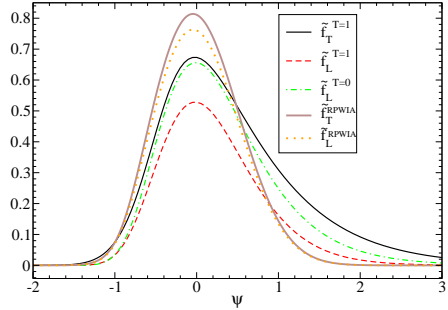}
  \caption[]{Reference scaling functions in the SuSAv2 model.}\label{fig1}
\end{figure}
In its original formulation the SuSA model for neutrino scattering relies on two hypotheses: 1) the longitudinal superscaling function is equal to the transverse one, $f_L=f_T\equiv f$, and 2) $f$ is the same in the isoscalar and isovector channels.

An improved version of the model (``SuSAv2'') has been recently elaborated by Gonzalez {\it et al.}~\cite{Gonzalez-Jimenez:2014eqa} to incorporate different effects arising in the microscopic RMF model in the longitudinal (L) and transverse (T) nuclear responses, as well as in the isovector (T=1) and isoscalar (T=0) channels. This has lead to the construction of three ``reference'' scaling functions, represented in Fig.~1 together with the corresponding functions in Relativistic Plane Wave Impulse Approximation (RPWIA), where there is no difference between the two isospin channels.
It appears that the SuSAv2 curves are lower and broader than the RPWIA ones, and display large high-energy (high $\psi$) tails, an effect of the strong final state interactions (FSI) of the model, as demonstrated in Ref.~\cite{Amaro:2006if}. 

The SuSAv2 scaling functions give an excellent representation of the quasielastic electron scattering data in a wide range of kinematics corresponding to medium and high momentum transfers, as extensively illustrated in Ref.~\cite{Gonzalez-Jimenez:2014eqa}. On the other hand, as expected, they fail to reproduce the data at low $q$, where nuclear collective modes become important and different theoretical schemes are required.

\begin{figure}[h]
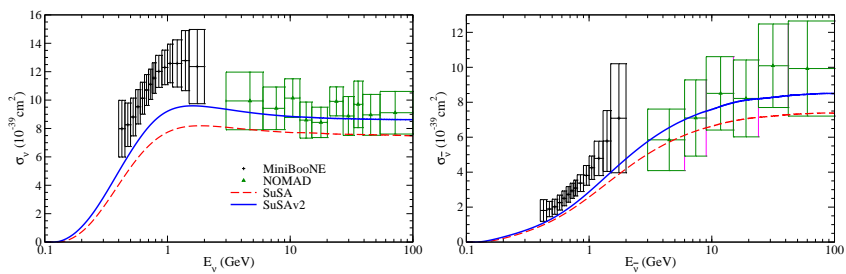

\centering
\includegraphics[scale=0.2]{Fig2a}
\includegraphics[scale=0.2]{Fig2b}
  \caption[]{CCQE $^{12}$C$(\nu,\mu^-)$ (left panel) and $(\bar{\nu},\mu^+)$ (right panel) cross section per nucleon presented as a function of the incident (anti)neutrino energy.
     Data from MiniBooNE~\cite{AguilarArevalo:2010zc,AguilarArevalo:2013hm} and NOMAD~\cite{Lyubushkin:2008pe} are compared with SuSA (dashed-red line) and SuSAv2 (solid-blue line) predictions.
     }\label{fig2}
\end{figure}

When applied to neutrino and antineutrino scattering, the agreement with the data depends on the kinematic conditions. This is illustrated in Fig.~2, where the SuSAv2 total quasielastic cross section on $^{12}$C (and for reference also the one associated to the simpler model, SuSA, evaluated in Ref.~\cite{Amaro:2013yna}) is displayed versus the (anti)neutrino energy and compared with the MiniBooNE~\cite{AguilarArevalo:2010zc,AguilarArevalo:2013hm} and NOMAD data~\cite{Lyubushkin:2008pe}. 
The SuSAv2 cross section is significantly
larger than SuSA one, as a consequence of the transverse enhancement of the model, although it still falls below the MiniBooNE data.
On the other hand both SuSA and SuSAv2 results are compatible 
with the NOMAD data, the latter being, in general, closer to the center of the
bins. 
 
The excess, at relatively low energy
($\langle E_\nu \rangle \sim0.7$ GeV), observed in MiniBooNE cross
sections has been interpreted as evidence that non-QE processes
may play an important role at that kinematics~\cite{Amaro:2010sd,
Martini:2011wp, Nieves:2011yp}. It is worth pointing out that in the
experimental context of MiniBooNE, ``quasielastic'' events are
defined as those from processes or channels containing no mesons in
the final state. Thus, in principle, in addition to the purely QE
process, which in this framework refers exclusively to processes induced
by one-body currents (IA), meson exchange current
effects (induced by two-body or many-body currents) should also be
taken into account for a proper interpretation of data.

In the NOMAD experiment the incident neutrino (antineutrino) beam
energy is much larger, with a flux extending from $E_\nu$= 3 to 100
GeV. In this case, one finds that data are in reasonable agreement
with predictions from impulse approximation models. 
Notice however that the large error bars of these data do not allow for further
definitive conclusions.

\begin{figure}[ht]
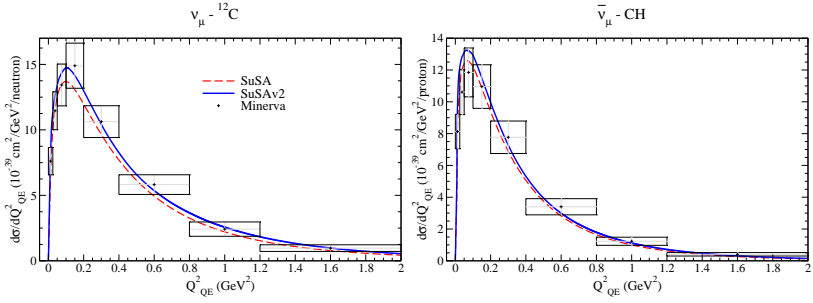

\centering
\includegraphics[scale=0.16]{Fig3a} %
\includegraphics[scale=0.16]{Fig3b}
  \caption[]{CCQE neutrino (left panel) and antineutrino (right panel) MINER$\nu$A data are compare with SuSA (dashed-red line)
       and SuSAv2 (solid-blue line) predictions.
     Data taken from~\cite{Minerva1,Minerva2}.}\label{fig3}
\end{figure}

In the MINER$\nu$A experiment the neutrino energy flux extends from
1.5 to 10 GeV and is peaked at $E_\nu\sim3$ GeV, {\it i.e.}, right in between the
MiniBooNE and NOMAD energy ranges. Therefore, its analysis can
provide very useful information on the role played by meson-exchange
currents in the nuclear dynamics.
In Fig.~3 the muon-neutrino (left) and antineutrino (right) single-differential CCQE cross sections ($d\sigma/dQ^2_{QE}$), measured by MINER$\nu$A~\cite{Minerva1,Minerva2}, are displayed as functions of the reconstructed four-momentum transfer squared, $Q^2_{QE}$, and compared with the SuSA and SuSAv2 results. In spite of the enhancement with respect to SuSA,
SuSAv2 is not only consistent, but it also improves the agreement
with MINER$\nu$A data. Furthermore, as expected, the SuSAv2 model produces very
close results to the RMF predictions, presented in
\cite{Megias:2014kia}. Thus, contrary to the MiniBooNE situation,
the comparison of MINER$\nu$A data and IA based models, in
particular, RMF and SuSAv2, leaves little room for MEC
contributions.

Before concluding this Section, we would like to comment on the microscopic origin of the superscaling function.
As already mentioned, the scaling function extracted from the data accounts for nuclear effects which
go beyond the RFG, among which NN correlations and FSI.
These, however, are difficult to be described in an unambiguous and
precise way and in most cases they are highly model-dependent. 
In order to shed light on the role played by these effects in inclusive lepton scattering, it is useful to formulate the problem using the language of spectral functions.
The spectral function $S(p,E)$ represents the probability to find a nucleon of momentum $p$ and energy $E$ in the nucleus. 
The RFG spectral function~\cite{CDM} is simply given by the product of an energy conserving $\delta$-function, indicating that nucleons are free and on-shell, and a $\theta$-function, indicating that the nuclear ground state is the filled Fermi sphere. This yields the well-known parabolic form of the superscaling function: $f_{RFG}(\psi)=\frac{3}{4}(1-\psi^2)$, very different from the experimental one~\cite{Jourdan}.

In Ref.~\cite{SF} a realistic spectral function $S(p,E)$
has been constructed that is in agreement with the experimental scaling
function. For this purpose effects of a finite energy spread have been included 
using natural orbitals (NO) for the single-particle
wave functions. Short-range NN correlations are accounted for within
the Jastrow correlation method.The results have been compared with the ones obtained using harmonic oscillator (HO) single particle wave functions.
Moreover FSI are accounted for by using an optical potential that leads
to an asymmetric scaling function, in accordance with the
experimental analysis, thus showing the essential role of the
FSI in the description of electron scattering reactions.
The results obtained using the above spectral functions
without FSI are in qualitative good agreement with those of
Refs.~\cite{Ben} and \cite{Ank}. 

\begin{figure}[ht]
\centering
\includegraphics[scale=0.23]{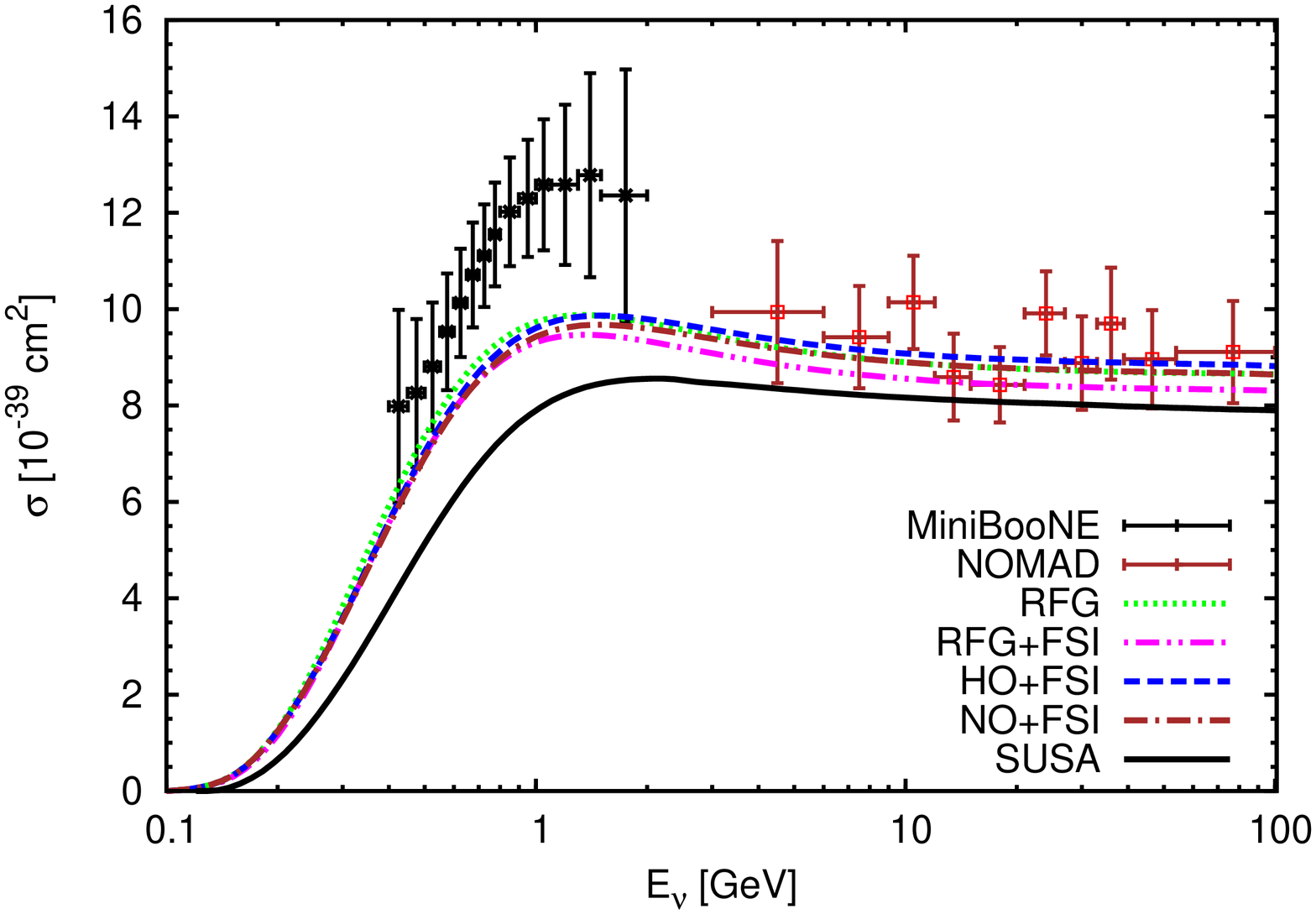}
\includegraphics[scale=0.23]{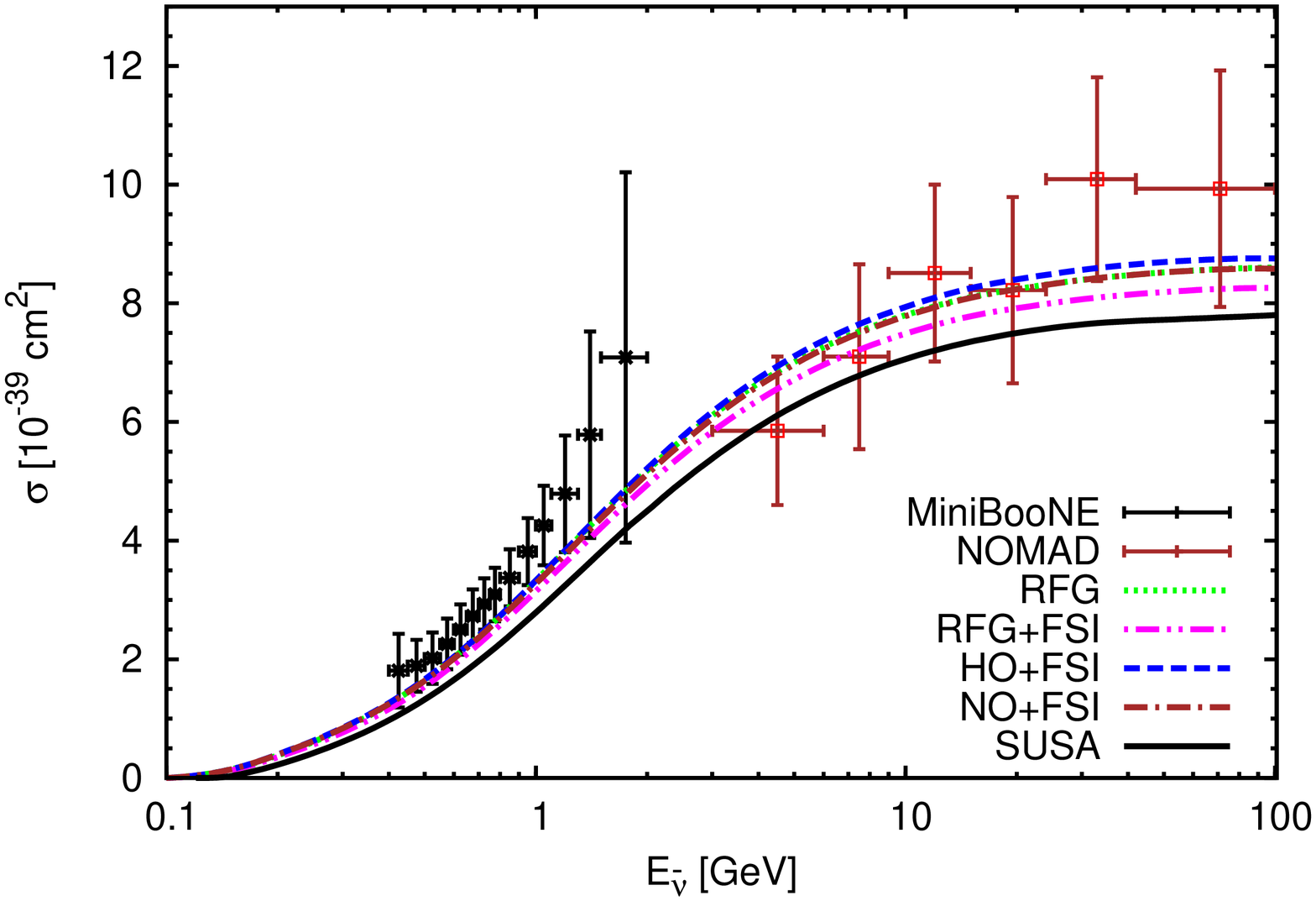}
  \caption[]{The data of Fig.~2 are compared with the prediction of the spectral function model described in the text, including NN correlations and final state interactions.}\label{fig4}
\end{figure}

In Fig.~\ref{fig4} the data already presented in Fig.~2 are compared with the results obtained within RFG+FSI, NO+FSI and HO+FSI approaches~\cite{Ivanov:2013saa}. As observed, all models give results very close to the SuSAv2 curves of Fig.~2, that agree with the NOMAD data but underpredict the MiniBooNE ones, more seriously in the neutrino than in the antineutrino case. 
It is also worth noticing that for very high $\nu_\mu$ ($\overline{\nu}_\mu$) energies the total cross section for neutrinos and antineutrinos is very similar. This is consistent with the negligible contribution given by the $T'$ (transverse-axial) response in this region. Only the $L$ and $T$ channels contribute for the higher values explored by NOMAD experiment (where the theory is in accordance with data). On the contrary, in the region explored by MiniBooNE, the main contributions come from the two transverse $T$, $T'$ channels, being constructive (destructive) in neutrino (antineutrino) cross sections. 
As already mentioned, effects beyond the IA, {\it i.e.,} $2p2h$ MEC, may have a significant contribution in the transverse responses leading to theoretical results closer to data. However, note that the enhancement needed to fit data should be larger for neutrinos than for antineutrinos, hence a careful analysis of $2p2h$ MEC contributions in both transverse responses is needed before more definitive conclusions can be drawn.

\section{Two-body currents}

Two-body currents correspond to the coupling of the virtual boson with a pair of interacting nucleons and can excite both one-particle-one-hole (1p1h) and two-particle-two-hole (2p2h) states. 
The corresponding Feynman diagrams can be classified into meson-exchange-currents (MEC), where the boson attaches to the meson exchanged between the two nucleons, and correlation currents, where it couples to one of the two nucleons. All diagrams must be considered in order to preserve current conservation.

In the RFG framework 1p1h excitations only contribute in the quasielastic peak region $-1<\psi<1$. Their effect is generally small in the vector sector, as shown in Ref.~\cite{PhysRep}, and can probably be neglected in first approximation, whereas a calculation including the axial current is still missing in the literature.

\begin{figure}[h]
\centering
\includegraphics[width=8cm, bb=130 280 410 780]{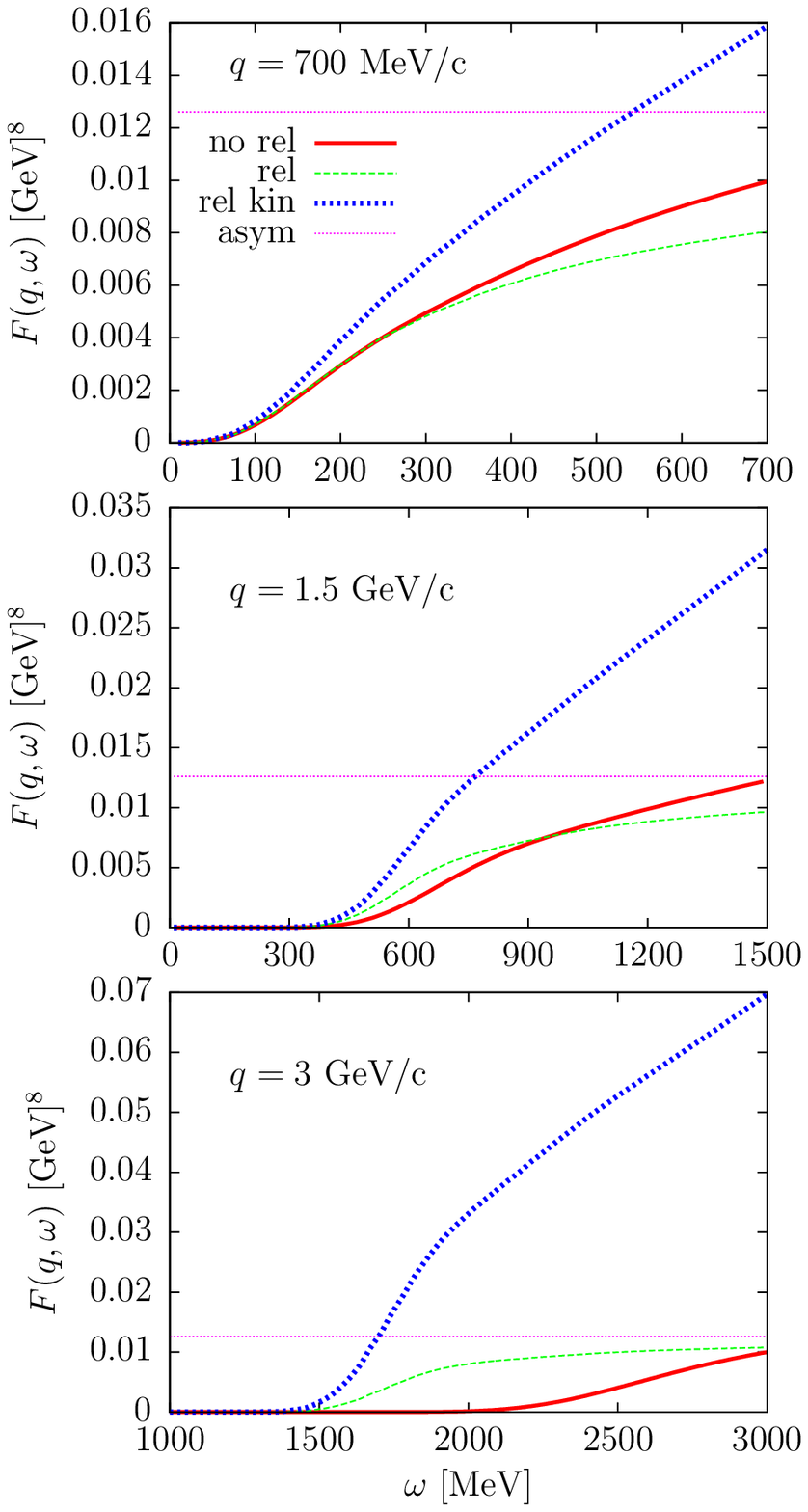}
  \caption[]{The 2p2h phase-space integral. Thick dotted lines: relativistic kinematics only without the relativistic factors $m_N/E$. Thin dashed lines: fully relativistic result.}\label{fig5}
\end{figure}

The 2p2h MEC on the contrary give a sizable contribution to the inclusive cross section~\cite{Martini:2011wp,Nieves:2011yp,Amaro:2010sd}.
The exact relativistic calculation, even in the simple RFG model, is computationally demanding, since it involves 7-dimensional integrals and some subtleties related to poles in the integrand function (see Refs.~\cite{Simo:2014wka,Simo:2014esa} for details). For the vector current, it has been performed by De Pace {\it et al.} in Ref.~\cite{DePace:2003xu} and the connection with scaling has been studied in Ref.~\cite{DePace:2004cr}. For the axial current, however, work is still in progress~\cite{SimoWIP}. Furthermore, in order to compare with neutrino cross section data, all the kinematics compatible with the experimental flux must be calculated: therefore it is necessary to design an optimal numerical procedure to reduce the computation time.

To this purpose we followed two different approaches. In Refs.~\cite{Amaro:2010sd,Amaro:2011aa} we used a parametrization of the exact calculation of De Pace {\it et al.}, which was performed for electron scattering at some kinematics and must be extrapolated to all kinematics involved in the neutrino flux folding integral. The corresponding neutrino and antineutrino cross sections have been compared with the MiniBooNE data and shown to be closer to the data than the pure one-body results, but still underpredicting the experimental point. Before drawing definitive conclusions one should however add the axial MEC to the model.

In parallel we have been revisiting the full exact calculation, including the axial current, trying to devise reliable approximations in order to optimize the numerical integration. 

To start with, in Refs.~\cite{Simo:2014wka,Simo:2014esa}
we have performed a careful analysis of the phase space.
We do not present the full analysis here for lack of space, but we just show one of the interesting outcomes of our study, pointing to the importance of a correct treatment of relativistic effects.

In Fig.~5 we show results for the 7-dimensional integral giving the phase-space function $F(q,\omega)$ at three values of the momentum transfer $q$. We compare the exact calculation (green online) with the non-relativistic result (red online) and a semi-relativistic result (blue online) obtained by implementing relativity in the kinematics but not in the current operators, which should include the 
appropriate Lorentz-contraction factors. It clearly emerges that
in order to ``relativize'' a non-relativistic 2p2h model,
implementing only relativistic kinematics is not only insufficient,
but it goes in the wrong direction: the effects coming solely from the 
relativistic kinematics lead to differences even larger than the discrepancy
between the non-relativistic and the fully relativistic
calculations.

\section{Summary and perspectives}

We have shown that the recently updated version of the Superscaling model (SuSAv2), based on the relativistic mean field model to account for the enhancement of the transverse response and for isospin effects, gives a good description of the high energy neutrino-nucleus data (NOMAD and MINER$\nu$A), while it underpredicts the MiniBooNE data. To possibly explain this discrepancy it is necessary to provide a reliable description of the meson exchange current contribution in the 2p2h sector. A fully relativistic and exact calculation of the associated response functions, involving both the vector and the axial two-body current, is under way and will be soon completed. Preliminary results point at the crucial importance of a consistent treatment of relativistic effects. 

\section*{Acknowledgements}
This contribution is based upon work partially supported by the Istituto Nazionale di Fisica Nucleare under Project ``MANYBODY''.

\end{document}